\documentstyle[aps,epsf]{revtex}
\def\prl#1#2#3{{ Phys. Rev. Lett.} {\bf #1}, #2 (#3)}
\def\ibid#1#2#3{{\it ibid.}  {\bf #1}, #2 (#3)}
\def\pla#1#2#3{Phys. Lett. A {\bf #1}, #2 (#3)}
\def\pre#1#2#3{Phys. Rev. E {\bf #1}, #2 (#3)}
\def\pra#1#2#3{Phys. Rev. A {\bf #1}, #2 (#3)}

\def\physd#1#2#3{Physica D {\bf #1}, #2 (#3)}

\def\etl{$et~al.$}

\def\beq{\begin{equation}}
\def\bc{\begin{center}}
\def\ec{\end{center}}
\def\eqn{\end{equation}}
\begin{document}

\title{Finite--time Lyapunov exponents of Strange Nonchaotic Attractors} 
\author{Awadhesh Prasad and Ramakrishna Ramaswamy} 
\address{School of Physical Sciences\\ Jawaharlal Nehru University, New
Delhi 110 067.}
\date{\today}
\maketitle
\begin{abstract} 
The probability distribution of finite--time Lyapunov exponents
provides an important characterization of dynamical attractors. We
study such distributions for strange nonchaotic attractors (SNAs)
created through several different mechanisms in quasiperiodically
forced nonlinear dynamical systems. Statistical properties of the
distributions such as the variance and the skewness also distinguish
between SNAs formed by different bifurcation routes.
\end{abstract}

\newpage
\section{INTRODUCTION}

The characterization of attractors in nonlinear dynamical systems is a
problem that has seen considerable progress in the past decade
\cite{bs,ott}. Beyond classification as simple or strange, the
calculation of fractal dimensions and the description of the measure in
term of a multifractal spectrum of singularities has become an
important means of describing the structure of dynamical attractors
\cite{bs}. For hyperbolic attractors there are rigorous results
connecting the multifractal structure, through the thermodynamic
formalism, with dynamical information as embodied in finite--time
Lyapunov exponents \cite{ott,grass}.

Grebogi \etl \cite{gopy} first described dynamical systems wherein the
attractors that result are fractal, but the dynamics is not chaotic,
in that the largest Lyapunov exponent is not greater than zero. These
strange nonchaotic attractors (SNAs) are generic in quasiperiodically
forced systems. Subsequently, considerable effort has been directed
toward the characterization and study of SNAs
\cite{rboag,kpf,chaos,hh,k,yl,lai}, which have
also been observed experimentally \cite{ditto,bulsara,newexp}. A
potential use of SNAs is in the area of secure communications, and
recent applications \cite{rr,chin} exploit the ease of synchronization
of such systems.

The strange nonchaotic state is only one of the possible dynamical
states realized in quasiperiodically driven systems; periodic,
quasiperiodic and chaotic attractors can also be obtained as parameters
are varied. SNAs are typically found for parameter values very close to
the boundaries of the chaotic regions, and the different bifurcation
mechanisms through which they are created is a problem of interest.
There are a number of routes or scenarios for the creation of SNAs,
some of which can be correlated with bifurcations. These include

(i) the Heagy--Hammel (HH) \cite{hh} mechanism involving a
collision between a period-doubled torus and its unstable parent,

(ii) the blowout bifurcation route \cite{lai},  and

(iii) intermittency \cite{prl}, when as a function of driving
parameter a chaotic strange attractor disappears and is eventually
replaced by a torus through an analogue of the
saddle-node bifurcation. 

The signature of these bifurcations in terms of the behaviour of the
largest Lyapunov exponent has been discussed in detail \cite{pre}.  The
blowout bifurcation mechanism \cite{lai} requires that the
quasiperiodic torus of a system with an invariant subspace losses its
transverse stability as a parameter changes across the transition and
lead to the birth of an SNA. In this process the transverse Lyapunov
exponent becomes positive while the nontrivial Lyapunov exponent for
the whole system remains negative.  In the HH mechanism \cite{hh}, as
system parameters are varied, the period-doubled torus gets
progressively more wrinkled and collides with a parent unstable torus;
this scenario is like an attractor merging crisis \cite{gory}.  The
distinctive signature of the intermittency route to SNA is a sharp
change in the Lyapunov exponent which shows large variance and scaling
behavior \cite{prl} at the bifurcation.

A general mechanism that is frequently observed but for which there is
no well-identified bifurcation is the so--called fractalization route
\cite{k}, whereby a smooth torus gets increasingly wrinkled and
transforms into a SNA without any interaction with a nearby unstable
periodic orbit (in contrast to HH). This is probably the most common
route to SNA in a number of maps and flows\cite{kpf,chaos}. 

The present paper addresses the issue of distinguishing among SNAs
formed by different routes through the use of finite--time (or local)
Lyapunov exponents. We show that the morphologies of different SNAs
differ in crucial ways, particularly for intermittent SNAs
\cite{prl,pre}. This is seen most dramatically in the characteristic
distributions of local Lyapunov exponents and the statistical
properties of the distributions such as the variance and the skewness.

In Sec.~II, we briefly introduce the dynamical systems that are studied
here. Results are discussed in Sec~III. This is followed by a summary
in Sec.~IV.

\section{DYNAMICAL SYSTEMS}

Several quasiperiodically driven systems---both maps and
flows\cite{kpf,chaos,hh,k,yl,lai}---have been shown to have SNAs. We
consider the quasiperiodically forced logistic map \cite{hh} wherein
three of the routes to SNAs can be observed. This system is
defined by the equations
\begin{eqnarray}
x_{n+1} &=& \alpha [1 + \epsilon \cos (2 \pi \phi_n)]~x_n~ ( 1 - x_n),
\nonumber \\ 
\phi_{n+1} &=& \phi_n + \omega ~~~~~({\rm mod ~} 1),
\label{lo}
\end{eqnarray}\noindent
where $x \in \hbox{R}^1$, $\phi \in \hbox{S}^1$, $\omega=
(\sqrt{5}-1)/2$ is the irrational driving frequency, and $\epsilon$
represents the forcing amplitude.

The largest nontrivial Lyapunov exponent $\Lambda$ is defined through
\begin{eqnarray}
\lambda_N (x_n) &=& \frac{1}{N} \sum_{j=1}^N \ln \vert dx_{n+j}/dx_{n+j-1}
\vert,\\ 
\Lambda &=& \lim_{N\to \infty} \lambda_N (x_n),
\label{lyap}
\end{eqnarray}\noindent
$\lambda_N$ being the local or $N$--step Lyapunov exponent. Note that
$\lambda_N$ depends on the initial condition $x_n$ while $\Lambda$,
of course, does not.

A region of the phase--diagram \cite{prl} of the forced logistic map,
Eq. (\ref{lo}) in the $\alpha$ -- $\epsilon^{\prime}$ plane ( where
$\epsilon^{\prime} =\epsilon/(4/\alpha -1 )$) is shown in Figure~1. The
rescaling of the forcing amplitude is done for convenience \cite{prl},
and the region of  Figure~1 corresponds to the period--3 window of
the unforced logistic map. The symbols P, S and C  correspond
to periodic, strange nonchaotic, and chaotic behavior of the system.
The dashed line is the locus of the period doubling bifurcation from
period--3 to period--6.

Along the left edge separating the periodic window from the region
C$_1$, there appear to be no SNAs.
On the right edge marked C$_1^{\prime}$, fractalized
SNAs are obtained. The intermittent SNAs are found in the region marked
{\bf I}.  Within this window, it was difficult to locate the HH
mechanism for the formation of SNAs which is known to operate in this
system for a number of different parameter values \cite{hh,prl}. 

We also study the blowout bifurcation route to SNA, which occurs in the
mapping \cite{lai} 
\begin{eqnarray}
x_{n+1} &=&  [a \cos(2 \pi \phi_n) + b] \sin(2 \pi x_n)
\nonumber \\ 
\phi_{n+1} &=& \phi_n + \omega ~~~~~({\rm mod ~} 1),
\label{lai}
\end{eqnarray}\noindent
($a$ and $b$ are parameters and $\omega= (\sqrt{5}-1)/2$) which is
bounded and has no other stable attractors other than the invariant
subspace $(x=0,\phi)$. As parameters changes across the critical
values, the dynamics of $x$ leads to a strange nonchaotic attractor
\cite{lai}. Our interest is in contrasting this mechanism for SNA 
formation which is also accompanied by on--off intermittency, with the
intermittent SNA \cite{prl}.

\section{RESULTS}

Although $\lambda_N$ depends on initial conditions, the probability
density, defined through
\beq
P(N, \lambda) d\lambda = {\rm Probability~ }~{\rm that~} \lambda_N {\rm
~lies~ between}~ \lambda ~ {\rm and}~ \lambda + d\lambda,
\eqn
does not. This distribution can be obtained by taking an (infinitely)
long, ergodic trajectory, and dividing it in segments of length $N$,
from which the local Lyapunov exponent can be calculated through Eq.~(2). 

For chaotic motion it has been argued \cite{ott,grass} that since the
local Lyapunov exponents can be treated as independent random
fluctuations, the central limit theorem is valid, leading to a normal
distribution for $\lambda_N$, 
\beq
P(N,\lambda) \approx \frac{1}{\sqrt{2\pi N G^{\prime
\prime}(\Lambda)}} \exp [- N G^{\prime \prime}(\Lambda)(\lambda
-\Lambda)^2/2] 
\eqn
with the function $G$, the spectrum of effective Lyapunov
exponents \cite{grass}, being appropriately defined \cite{ott}.

These expectations are not always satisfied since there can be
important correlations in the dynamics. We have recently described the
characteristic distributions for finite time Lyapunov exponents in low
dimensional chaotic systems where there are significant departures
from central--limit behaviour \cite{pr}.

For SNAs there are additional complications. Although $\Lambda$ is by
definition nonpositive, $P(N,\lambda)$ can have a significant
contribution from $\lambda > 0$: for some of the time, these systems
behave chaotically because of the fractal structure of the
attractors\cite{chaos}.  In the limit of large $N$, the contribution
from positive $\lambda$ decreases and the density collapses to a
$\delta$--function, $\lim_{N \to \infty} P(N,\lambda) \to
\delta(\Lambda-\lambda)$. 

Shown in Figs.~2 and 3 are local Lyapunov distributions for the four
routes to SNAs (parameters are specified in the caption), for short ($N
= 50$) and long but finite ($N = 1000$) times. The fractalized and HH
SNAs both show a gradual approach to the normal distribution (a
gaussian is fit to the data in Fig.~3), while the blowout SNAs, and
more spectacularly, the intermittent SNAs show a distinctive departure
from the gaussian distribution.

The intermittent SNA is morphologically and dynamically very different
from the other SNAs, and the shape of the characteristic distribution,
$P(N, \lambda)$ is a combination of a gaussian and an exponential
\cite{pr}. In contrast to the other SNAs, the distribution is
asymmetric, and the large $\lambda$ tail decays very slowly.

It appears that the intermittent SNA is in a distinct universality
class \cite{pr} and a number of quantitative measures can be devised in
order to show this distinction.  Consider, for example, the fraction of
exponents lying above $\lambda=0$,
\beq
F_+(N) = \int_0^{\infty} ~P(N,\lambda)~ d\lambda,
\eqn
$F_+(N)$ vs $N$ for the different SNAs are shown in Fig.~4. Except for
the intermittent SNA, for which $F_+(N) \sim N^{-\beta},$ this quantity
decays exponentially, $ F_+(N) \sim \exp(-\gamma N)$, with the
exponents $\beta$ and $\gamma$ depending strongly on the parameters of
the system. We found that the values of the exponents are $\beta =
0.72$ for the intermittent SNA and $\gamma = 0.02, 0.007$ and $0.042$
respectively for HH, fractalized, and blowout SNAs. Similarly, other
statistical properties of the these distributions can be studied.  We
calculate the first two moments about the arithmetic mean of all
distributions and obtain the variance,
\beq 
\sigma^2=
\int_{-\infty}^{\infty} (\lambda-\Lambda)^2~P(N,\lambda)~ d\lambda,
\eqn
and the coefficient of skewness, namely 
\beq
\gamma_1= \int_{-\infty}^{\infty} (\lambda-\Lambda)^3 ~P(N,\lambda)~
d\lambda/(\sigma^3),
\eqn
which are shown in Fig.~5(a) and Fig.~5(b) respectively. Generally for
all types of SNAs, the variance of $P(N,\lambda)$ decreases as a power
of $N$, $\sigma^2 \sim 1/N^{\delta}$ where the exponent $\delta$ is
different for each SNA. The variance for intermittent SNAs decreases
very slowly compared to other SNAs, and our numerical results for the
exponents, for the examples shown here are $ = 0.97, 1.71,
1.63,\mbox{~and~} 1.7 $ for intermittent, fractalized, HH, and blowout
SNAs. The degree of asymmetry in the distribution is quantified by the
significantly larger skewness $\gamma_1$ (see Fig.~5(b)).

\section{SUMMARY}

In the present paper we have studied the dynamical structure of strange
nonchaotic attractors formed by different bifurcation mechanisms in
quasiperiodically driven systems, by examining the {\it distribution}
of finite--time Lyapunov exponents. Although the Lyapunov exponent is
negative on a SNA, over short times, nearby trajectories can separate
from one another since the attractor is strange: this corresponds to a
local positive Lyapunov exponent. The manner in which this distribution
changes as a function of time is characteristic of the attractor, and
of the bifurcation routes through which attractors are created:
intermittent dynamics leads to very distinctive distributions of local
Lyapunov exponents \cite{pr}.

Our present results further underscore the utility of finite--time
Lyapunov exponents in describing the local structure of dynamical
attractors \cite{ott,grass,pr} in general. For the case of hyperbolic
attractors, the theory connecting these to the invariant measure is
well--developed. The present paper is part of a preliminary
step towards understanding the connection between an underlying fractal
structure and globally nonchaotic dynamics on strange nonchaotic
attractors. 

\vskip1cm
\centerline{\sc ACKNOWLEDGMENT} This research was supported by 
grant No. SP/S2/E07/96 from the Department of Science and Technology, India. We thank
Vishal Mehra for discussions.


\newpage
\begin{thebibliography}{99}

\bibitem{bs} C. Beck and F. Schl\"ogl, {\it Thermodynamics of Chaotic
systems}, (Cambridge University Press, Cambridge, 1993)

\bibitem{ott} E. Ott, {\it Chaos in dynamical systems }, (Cambridge
University Press, Cambridge, 1994).

\bibitem{grass} P. Grassberger, R. Badii, and A. Politi, J. Stat. Phys.
{\bf 51}, 135 (1988).

\bibitem{gopy} C. Grebogi, E. Ott, S. Pelikan, and J.A. Yorke,
\physd{13}{261}{1984}.


\bibitem{rboag} F.J. Romeiras, A. Bondeson, E. Ott, T.M. Antonsen,
and C. Grebogi, \physd{26}{277}{1987}.

\bibitem{kpf} T. Kapitaniak, E. Ponce, and J. Wojewoda,
\pla{154}{249}{1991}; S.P. Kuznetsov, A.S. Pikovsky, and U. Feudel,
\pre{51}{R1629}{1995};  U.  Feudel, J. Kurths and A.
Pikovsky,\physd{88}{176}{1995}; O. Sosnovtseva, U. Feudel, J. Kurths,
and A.  Pikovsky, \pla{218}{255}{1996}; A. Venkatesan and M.
Lakshmanan, \pre{55}{5134}{1997}.

\bibitem{chaos} A. Pikovsky and U. Feudel, Chaos, {\bf5}, 253 (1995).

\bibitem{hh} J.F. Heagy and S.M. Hammel, \physd{70}{140}{1994}.

\bibitem{k} K. Kaneko, Prog. Theor. Phys., {\bf 71}, 1112 (1984); T.
Nishikawa and K. Kaneko, \pre{54}{6114}{1996}. 

\bibitem{yl} T. Yal\c{c}inkaya and Y.C. Lai, \prl{77}{5039}{1996},
\pre{56}{1623}{1997}. 

\bibitem{lai} Y.C. Lai, \pre{53}{57}{1996}; Y.C. Lai, U. Feudel, and C.
Grebogi, \ibid{54}{6070}{1996}.

\bibitem{ditto} W. L. Ditto, M. L. Spano, H. T. Savage, S. N. Rauseo,
J. Heagy, and E. Ott, \prl{65}{533}{1990}.

\bibitem{bulsara} T. Zhou, F. Moss, and A. Bulsara, \pra{45}{5394}{1992}.

\bibitem{newexp} W.X. Ding, H. Deutsch, A. Dinklage, and C. Wilke,
\pre{55}{3769}{1997}.

\bibitem{rr} R. Ramaswamy, \pre{56}{7294--96}{1997}

\bibitem{chin} C. Zhou and T. Chen, Europhys. Lett., {\bf 38}, 261 (1997).

\bibitem{prl} A. Prasad, V. Mehra, and R. Ramaswamy,\prl{79}{4127}{1997}.

\bibitem{pre}A. Prasad, V. Mehra and R. Ramaswamy, Phys. Rev. E, in press (1998).

\bibitem{gory} C. Grebogi, E. Ott, F.J. Romeiras, and J.A. Yorke,
\pra{36}{5365}{1987}; V. Mehra and R. Ramaswamy, \pre{53}{3420}{1996}. 

\bibitem{pr} A. Prasad and R. Ramaswamy, to be published.

\end{thebibliography}
\end{document}